\documentclass{aa}
\usepackage{graphicx}
\usepackage{txfonts}
\begin{document}
   \title{Checking the reliability of equivalent width $R_{23}$
  for estimating the metallicities of galaxies }

  \titlerunning{Checking the EW $R_{23}$ method}

   \author{Y. C. Liang\inst{1,2}, F. Hammer\inst{2}, and S. Y. Yin\inst{1,3,4}}

   \authorrunning{Liang et al.}

   \offprints{Y. C. Liang,
              email: ycliang@bao.ac.cn
          }
  \institute{$^1$National Astronomical Observatories, Chinese Academy of Sciences,
 20A Datun Road, Chaoyang District, Beijing 100012, China \\
$^2$GEPI, Observatoire de Paris-Meudon, 92195 Meudon, France  \\
$^3$Department of Physics, Hebei Normal University, Shijiazhuang 050016, China \\
$^4$Department of Physics, Harbin University, Haerbin 150086, China
}

   \date{Received; accepted 8 August 2007}
  \abstract
   {}
   {We verify whether the O/H abundances of galaxies can be derived from
the equivalent width (EW) $R_{23}$ instead of the extinction-corrected
flux $R_{23}$, and eventually propose a method of improving the reliability
of this empirical method, which is often used for the non-flux
calibrated spectra of galaxies.}
   {We select 37,173 star-forming galaxies from the Second Data
Release of the Sloan Digital Sky Survey (SDSS-DR2), which offers a wide
range of properties to test the EW method. }
   {The EW-$R_{23}$ method brings with it a significant bias: for the bulk
of SDSS galaxies, it may affect the determination of log(O/H) by factors
ranging from -0.2 to 0.1\,dex and for some galaxies by factors ranging
from -0.5 to 0.2\,dex. We characterize this discrepancy (or bias) by
$\alpha$ = $\rm ({I_{[OII]}/I_{H\beta}}$)/($\rm {EW_{[OII]}/EW_{H\beta}})$,
which is virtually independent
of dust extinction, while tightly correlating with $D_n$(4000),
although at a lower significance, with ($g-r$) colors. }
   {The EW-$R_{23}$ method cannot be used as a proxy for the
extinction-corrected flux $R_{23}$ method. From analytical third-order
polynomial fits of $\alpha$ versus ($g-r$) colors,
we have been able to correct
the EW-$R_{23}$ method. With this additional and easy correction,
the EW-$R_{23}$
method provides O/H abundance values similar to those derived from the
extinction-corrected flux $R_{23}$ method with an accuracy of $\pm$0.1\,dex
for $>$92\% of the SDSS galaxies.
   }
   \maketitle

\section {Introduction}

Chemical abundance is a fundamental parameter for
tracing the history of star formation and evolution of galaxies.
Oxygen is the most commonly used metallicity
indicator in the interstellar medium (ISM) by virtue of its high relative
abundance and strong emission lines in the optical part of the spectrum
(e.g., [O~{\sc ii}]$\lambda $3727 and [O~{\sc iii}]$\lambda\lambda $4959, 5007).
However, the ``direct" method of estimating oxygen abundances from
electron temperature ($T_e$) is generally only available for
metal-poor galaxies (12+log(O/H)$<$8.5),
where the [O~{\sc iii}]$\lambda $4363 emission
line can possibly be detected, which
is needed for measuring $T_e$ by its ratio
to [O~{\sc iii}]$\lambda\lambda $4959, 5007
(Pagel et al. 1992; Skillman \& Kennicutt 1993).
For metal-rich galaxies,
the most commonly used are empirical strong-line methods,
such as $R_{23}$:
\begin{equation}
R_{23} = \frac{I([\rm O II]\lambda3727) +
I([{\rm O III}]\lambda\lambda4959,5007)}{I(\rm H\beta)},
\label{eq1}
\end{equation}
i.e. the flux ratios of [O~{\sc ii}] and [O~{\sc iii}] to H$\beta$
(Pagel et al. 1979; McGaugh 1991; Zaritsky et al. 1994;
Kobulnicky et al. 1999;
Tremonti et al. 2004 and the references therein).

However, flux calibrations are frequently problematic in the current
generation of wide-field galaxy surveys of multiobject spectrography,
because of unfavorable observing conditions or instrumental effects such
as a variation in system response over the field of view, nevertheless, one
still expects to derive metallicity properties of the
star-forming galaxies detected in the large data sets from surveys.
Then, $equivalent~~ widths$ (EWs) are being used to
replace the fluxes of their $R_{23}$ values for
estimating metallicities of the galaxies, i.e., from
the EW\,$R_{23}$:
\begin{equation}
 \rm EW~R_{23} = \frac{EW([\rm O II]\lambda3727) +
 EW([{\rm O III}]\lambda\lambda4959,5007)}{EW(\rm H\beta)}.
\label{eq2}
\end{equation}
This replacement was first checked by
Kobulnicky \& Phillips (2003; KP03 hereafter)
on the basis of a sample of 243 nearby galaxies.
Consequently, this method has been used by some researchers
to estimate metallicities of nearby and even intermediate-$z$
 galaxies (e.g. Kobulnicky et al. 2003;
Kobulnicky \& Kewley 2004; Lamareille et al. 2005a,b; Mouhcine et al. 2006a,b
etc.).

However, it is known that
there is a continuum ($F_{C\lambda}$) scale factor
between the flux ($F_{\lambda}$) and EW ($W_{\lambda}$) values of
the emission line:
\begin{equation}
W_\lambda = \frac {F_{\lambda}} {F_{C\lambda}}.
\end{equation}
This will naturally make us ask
whether the continua underlying [O~{\sc ii}] and
H$\beta$ (and [O~{\sc iii}]) are very similar or not.
If not, $F_{C\lambda}$ could be far from a unique constant,
and then the metallicities derived from EW\,$R_{23}$
could have large discrepancies from those derived from flux $R_{23}$.

The SDSS provides a complete dataset and measurements
up to several ten-thousand galaxies,
making it a very good database for studying this question.
In this paper, we selected 
37,173 star-forming galaxies
from the SDSS-DR2 to further check the reliability of
using EW\,$R_{23}$ replacing the
extinction-corrected flux $R_{23}$
to estimate the metallicities of galaxies.
Moreover,
this is a large homogeneous database observed by one single highly efficient
facility, which will minimize the effect of using various instruments.
Some other characteristic parameters provided by the
SDSS can also help for understanding the question further, e.g.,
$D_n$(4000), $g$ and $r$ photometric magnitudes, etc.

This paper is organized as follows. The sample selection criteria are
described in Sect.2. In Sect.3, we check how big the difference is between
the underlying continua of [O~{\sc ii}] and H$\beta$,
which is quantified as a parameter $\alpha$ (by 1/$\alpha$).
The emission-line quantities of the sample galaxies
are analyzed in Sect.4, as well as
the discrepancies between their EW\,$R_{23}$ and flux $R_{23}$,
and the discrepancies between the derived
log(O/H) abundances from them.
In Sect.5, we try to find the factors that mostly affected 
the $\alpha$ parameter,
then to find the relations between them and $\alpha$,
hence to modify the EW\,$R_{23}$ method,
which includes the stellar population
indicators $D_n$(4000) and colors.
In Sect.6, we discuss the boundary of the upper branch of 
12+log(O/H) abundances from the log$R_{23}$ calibration,
then summarize and conclude this paper in Sect.7.

\section{Sample selection}
\label{sect.2}

\begin{figure}
\centering
\input epsf
\epsfverbosetrue
\epsfxsize 7.2cm
\includegraphics[bb=21 149 573 575,width=7.2cm,clip]{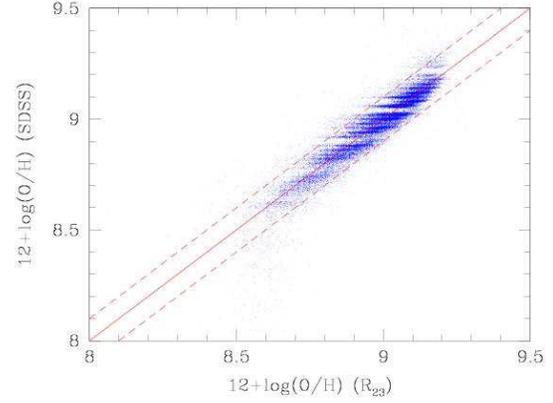}
\caption{Comparison between the oxygen abundances
obtained by the MPA/JHU group (marked as `SDSS') and those derived from
$R_{23}$ calibration of Tremonti et al. (2004) (marked as `$R_{23}$').
 The dashed lines mark the 0.1\,dex
discrepancies from the equal values (solid line).  
       }
\label{figoh}
\end{figure}

The data analyzed in this study were drawn from the
SDSS-DR2 (Abazajian et al.
2004). These galaxies are part of the SDSS ``main" galaxy sample 
used for large-scale structure studies (Strauss et al. 2002).
We selected a sample of star-forming galaxies with metallicity
estimates from the SDSS-DR2 database.  The selection criteria are
similar to those of Tremonti et al. (2004) and Liang et al.
(2006). We summarize the criteria for selection as follows,
and mark the selected number of the sample galaxies in 
parenthesis at the end of each step:
\newline
(i) SDSS-DR2 (261,054), 14.5$<$r$<$17.77 mag (Petrosian magnitude, 193,890 left);
\newline
(ii) 12+log(O/H)$_{\rm SDSS}$ $>$ 0 (50,385 left) (SDSS refers to
the metallicity values
provided by the MPA/JHU group, which were obtained on the basis of
the 2001 Charlot \& Longhetti model and Bayesian technique,
see Tremonti et al. 2004);
\newline
(iii) $0.04<z<0.25$ (40,693 left) (this allows the spectral coverage from
[O~\textsc{ii}]$\lambda$3727 to [S~\textsc{ii}]$\lambda$6731 for the sample
galaxies);
\newline
(iv) the emission lines
[O~\textsc{ii}], H$\beta $, [O~\textsc{iii}], H$\alpha $, and [N~\textsc{ii}] are
detected and their fluxes and EWs are reasonable;
 fluxes of emission lines H$\beta $, H$\alpha $, and [N~\textsc{ii}]6584
are detected at levels higher than 5$\sigma $ (39,029 left);
\newline
(v) 8.5$<$12+log(O/H)$_{R_{23}}$$<$9.3 (38,932 left) (we
adopt the formula of Tremonti et al. 2004 to convert $R_{23}$ to
the oxygen abundance 12+log(O/H)$_{R_{23}}$, which is only
suitable for metal-rich galaxies; therefore, we only select the
galaxies with 12+log(O/H)$>$8.5 here;  9.3 is almost the upper
limit of the abundances in the samples);
\newline
(vi) the discrepancy between log(O/H)$_{R_{23}}$
and log(O/H)$_{\rm SDSS}$ is less than 0.1\thinspace dex
(37,173 left finally), which removes some scattered data points
from the $R_{23}$ calibration, as shown by Fig.~\ref{figoh}.

Finally, we obtained a sample of 
37,173 
star-forming galaxies from the SDSS-DR2 as our sample in
this study.

The criteria (i)-(iv) are almost the same as what was adopted in 
Tremonti et al. (2004),
except the lower limit for redshift $z$ was increased from 0.005 to 0.04 by following 
Kewley et al. (2005) to minimize the aperture effects of the SDSS.
In Sect.~\ref{sec60}, we discuss criterion (v) in particular,
i.e. 12+log(O/H)$\sim$8.5 as 
the lower boundary of the upper branch of oxygen abundances
from log$R_{23}$ calibration.

\section{Quantifying the difference between
 EW\,$R_{23}$ and flux $R_{23}$: determining the $\alpha$ parameter }
\label{sect.3}

The MPA/JHU collaboration has put the measurements of emission-lines
and some physical parameters for a large sample of SDSS galaxies on
the MPA SDSS website\footnote{http://www.mpa-garching.mpg.de/SDSS/}
(Kauffmann et al. 2003; Brinchmann et al. 2004; Tremonti et al. 2004 etc.).
These measurement values were obtained
from the stellar-feature subtracted spectra
with the spectral population synthesis code of Bruzual \& Charlot (2003).

 Fluxes of the emission lines should be corrected for dust extinction,
so we estimate the attenuation of the sample galaxies
using the Balmer line ratio
H$\alpha$/H$\beta$ and assuming case B recombination, with a density of 100
cm$^{-3}$, a temperature of $10^4$ K, and the intrinsic
H$\alpha$/H$\beta$ ratio of 2.86 (Osterbrock 1989),
following the relation of
\begin{equation}
(\frac{I_{H\alpha}}{I_{H\beta}})_{obs}=
(\frac{I_{H\alpha0}}{I_{H\beta0}})_{intr}$10$^{-c(f(H\alpha)- f(H\beta))}.
\end{equation}
Using the average
interstellar extinction law given by Osterbrock(1989), we obtain
$f$(H$\alpha$)- $f$(H$\beta$) = $-$0.37.
For
the data points with $c < 0$, $c$ = 0 is assumed since their
intrinsic H$\alpha$/H$\beta$ may be lower than 2.86 if their
electron temperature is high (Osterbrock 1989).

Then, the extinction-corrected $R_{23}$ parameter and its relation
with the EW\,$R_{23}$ are:
\begin{eqnarray}
 & &  R_{23}  = {  {I_{\rm [OII]} + I_{\rm [OIII]}} \over {I_{\rm H\beta}} } \nonumber \\
 & = & { {F_{\rm [OII]}} \over {F_{H\beta}} }10^{c({f(\rm [OII])}-f(\rm H\beta))}
  + { {F_{\rm [OIII]}} \over {F_{\rm H\beta}} } 10^{c{(f(\rm [OIII])}-f(\rm H\beta))} \nonumber  \\
 & = &   { {W_{\rm [OII]}F_{\rm C,[OII]}} \over {W_{\rm H\beta}F_{\rm C,H\beta}}
  }10^{c{(f(\rm [OII])}-f(\rm H\beta))}  \nonumber  \\
 &   & +  { {W_{\rm [OIII]}F_{\rm C,[OIII]}} \over {W_{\rm H\beta}} F_{\rm
 C,H\beta}} 10^{c{(f(\rm [OIII])}-f(\rm H\beta))}     \nonumber  \\
   & = & {{ \alpha {W_{\rm [OII]}} + {W_{\rm [OIII]}}} \over {W_{\rm H\beta}}},
 \label{R23}
 \end{eqnarray}
where
$\alpha = ({F_{\rm C,[OII]}}/F_{\rm C,H\beta}) 10^{c{(f(\rm
[OII])}-f(H\beta))}$, and
$({F_{\rm C,[OIII]}}/F_{\rm C,H\beta}) 10^{c{(f(\rm
[OIII])}-f(H\beta))}$ is about equal to 1 since
[O~{\sc iii}]$\lambda\lambda $4959,5007 and H$\beta$ 
are very close in wavelength.
The expression $I_{\rm [OII],[OIII],H\beta}$  refers to
the dereddened, calibrated flux values of the corresponding lines;
$F_{\rm [OII],[OIII],H\beta}$ refers to the observed flux values;
and
$W_{\rm [OII],[OIII],H\beta}$ represents the EW values of the
related emission lines. The parameter $c$ is the extinction coefficient.
(Also see KP03)

The $\alpha$ values of these SDSS galaxies can be calculated directly
from
$\alpha = ({F_{\rm C,[OII]}}/F_{\rm C,H\beta})
10^{c{(f(\rm [OII])}-f(H\beta))}$,
where $F_{\rm C,[OII]}$ and $F_{\rm C,H\beta}$ can be estimated from
their ratios of fluxes to EW values, and $c$ is the extinction coefficient.
The derived
$\alpha$ parameters of these sample galaxies
show a median value of 0.85
and an average value of 0.86 in a range from 0.1 to 2.6.
KP03 estimate the typical value of the
$\alpha$ parameters of their 243 sample galaxies to be $\alpha = 0.84 \pm 0.3$.
However, KP03 still adopted $\alpha=1$ when they used the EW\,$R_{23}$
to estimate the oxygen abundances of galaxies.

\section{Analysis of emission-line quantities }

 Several relations are analyzed in this section, including
 those of the emission-line quantities, e.g. EW(H$\beta$),
 with the $\alpha$ parameter,
 with the discrepancies of EW\,$R_{23}$ and flux $R_{23}$,
 and with the discrepancies of the log(O/H) abundances
 derived from these two $R_{23}$ values.

\subsection{Relations between the line strengths and $\alpha$ parameters}

Figure~\ref{fig1}a shows the relations between
the emission-line strengths EW(H$\beta$) and the line-ratio
$\rm {EW_{[OII]} \over EW_{H\beta}}$/$\rm {I_{[OII]}\over I_{H\beta}}$
(=1/$\alpha$) of the sample galaxies. The solid line marks the one-to-one
correspondence.
It seems that there is some correlation between the $\alpha$
parameter and the emission-line strengths:
from the galaxies with stronger emission
lines, the ratios of
log($\rm {EW_{[OII]} \over EW_{H\beta}}$/$\rm {I_{[OII]}\over I_{H\beta}}$)
(=log(1/$\alpha$)) change from -0.2 (mostly) to 0.5 monotonicly,
though the scatters are large and show some exceptional points
with log(1/$\alpha$)$<$-0.2.
This range is similar to what KP03 found for their sample galaxies.
The line strengths of [O~{\sc ii}] and [O~{\sc iii}],
i.e., EW$_{\rm [OII]}$ and EW$_{\rm [OIII]}$,
also show similar trends to EW(H$\beta$)
in these kinds of relations, but with somewhat larger scatter.
 The ratio
log($\rm {EW_{[OII]} \over EW_{H\beta}}$/$\rm {I_{[OII]}\over I_{H\beta}}$)
(=log(1/$\alpha$)) of all the sample galaxies show 
a median value of 
0.069\,dex 
(with a scatter of 
0.060\,dex)
and a mean value of 
0.077\,dex 
(with a scatter of 
0.104\,dex).

\subsection{Discrepancies between EW $R_{23}$ and $R_{23}$}

When directly using the EW\,$R_{23}$ to replace the
flux $R_{23}$, namely,
$\alpha$=1 is adopted for EW\,$R_{23}$,
the derived metallicities could have some discrepancies.
 Figure~\ref{fig1}b shows the discrepancies between the
EW\,$R_{23}$ and the extinction-corrected flux $R_{23}$ as a function of
line strengths EW(H$\beta$).
 The general trend is that,
 from galaxies with stronger line strengths,
  the differences between log(EW\,$R_{23}$) and
 log($R_{23}$) increase from -0.15 (mostly) to 0.5\,dex monotonicly.
KP03 also find a similar trend for their sample galaxies.
Our much larger sample
shows this systematic discrepancy more clearly.
 The median value of these discrepancies is about  
 0.061\,dex (with a scatter of 
 0.050\,dex),
 and the mean value of them is about 
 0.069\,dex (with a scatter of
 0.086\,dex). 

\subsection{Discrepancies between (O/H)$_{EWR_{23}}$ and (O/H)$_{R_{23}}$}

\begin{figure}
\centering
\input epsf
\epsfverbosetrue
\epsfxsize 7.2cm
\includegraphics[bb=110 283 249 497,width=7.2cm,clip]{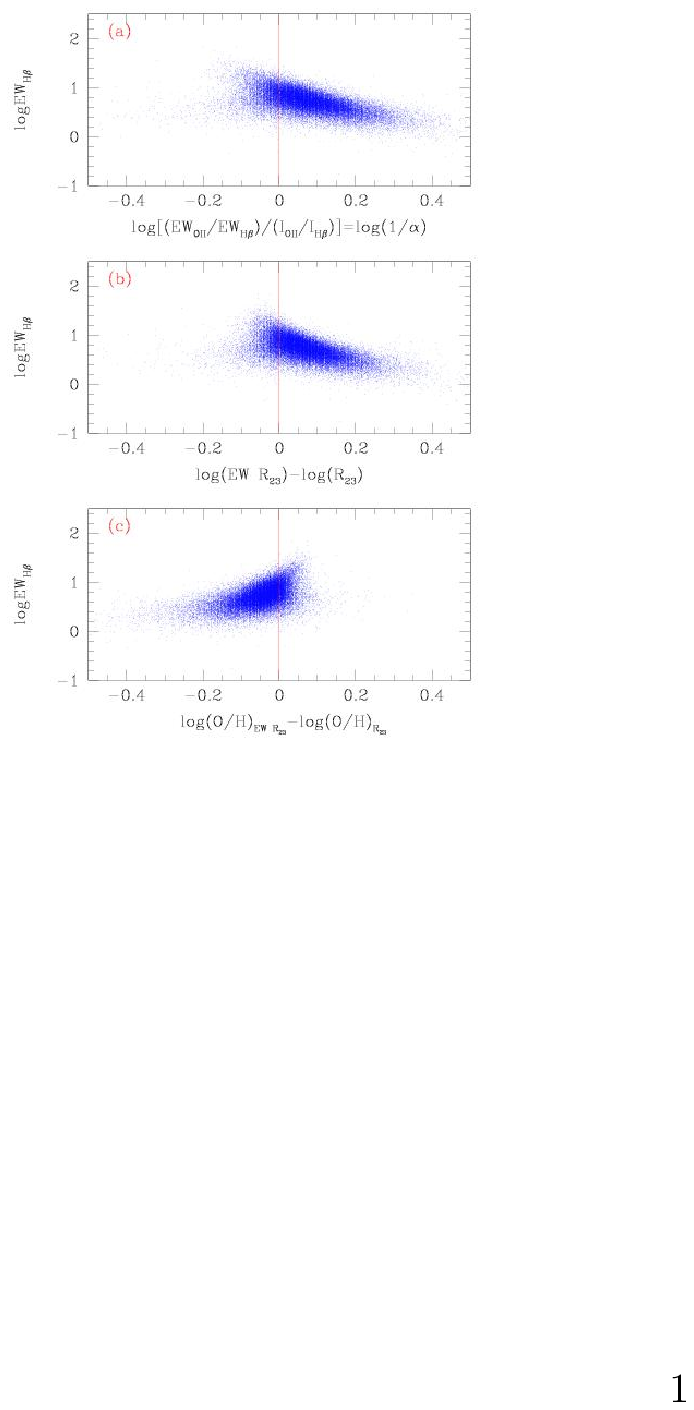}
\caption{ {\bf (a).} Emission-line
strengths EW(H$\beta$) of the sample galaxies as
a function of the ratios of equivalent width ratios to
dereddened emission-line fluxes ratios for [O~{\sc ii}]/H$\beta$.
       {\bf (b).} EW(H$\beta$) as a function of
       the discrepancy between
       the quantity of EW\,$R_{23}$ and the
       flux $R_{23}$ (extinction-corrected).
      {\bf (c).} EW(H$\beta$) as a function of
      the discrepancy between the two metallicity
      estimates from EW\,$R_{23}$ and flux $R_{23}$.
       }
\label{fig1}
\end{figure}

 One of the most important results of this work is given in
 Fig.~\ref{fig1}c, which shows the difference between
 log(O/H)$_{EW~R_{23}}$ and log(O/H)$_{R_{23}}$
  as functions of the
 emission-line strengths EW(H$\beta$).
 The general trend is that,
 from the galaxies with stronger line strengths,
 the differences in the two log(O/H) estimates change from -0.5 to 0.2\,dex,
 with most of them ranging from -0.2 to 0.1\,dex. KP03 do not present such
 a direct comparison for the O/H abundances.

 The large sample of SDSS star-forming galaxies that we use
 generally show that
 the EW\,$R_{23}$ will underestimate the oxygen abundances of the
 sample galaxies by a factor of
  0.041 
  (the median offset, with a scatter of 
  0.036\,dex) or
  0.054\,dex  (the mean offset, with a scatter of 
  0.078\,dex). 
 Here the log(O/H) abundances were
 obtained from the $R_{23}$ calibration
 of Tremonti et al. (2004).
 We also adopted some other $R_{23}$  calibration formulas,
 i.e. Kobulnicky et al. (1999, the analysis formulas for the models of
 McGaugh 1991), Zaritsky et al. (1994), and
 Kobulnicky \& Kewley (2004, the average of McGaugh 1991 and
 Kewley \& Dopita 2002),
  to check these discrepancies, the results of which are quite similar.
  Moustakas \& Kennicutt (2006) find similar
 discrepancies to ours by comparing the
 abundances derived from the extinction-corrected flux- and EW-$R_{23}$
 of 12 nearby spiral galaxies.
 They find that the integrated abundances determined from the
 emission-lines systematically underestimated the characteristic abundances.
 The mean offset is $-$0.06$\pm$0.09\,dex using the McGaugh (1991)
 calibration, or $-$0.11$\pm$0.13\,dex using the Pilyugin \& Thuran (2005)
 calibration, and the corresponding
 median offsets are $-$0.04 and $-$0.09\,dex, respectively.

  These results show that
   the oxygen abundances derived from the
  EW\,$R_{23}$ and the extinction-corrected flux-$R_{23}$ are not seriously
  different for these star-forming galaxies, generally less than 0.1\,dex.
  However, the global discrepancy of them
  does show a wide distribution, from -0.5 to 0.2\,dex,
  which means that this replacement of EW\,$R_{23}$ to flux $R_{23}$
  for metallicity estimates
  could cause different
  effects on the individual galaxies, and should be
  considered carefully.

\section{Modifying the EW $R_{23}$-method}

   It is interesting to further check
   the main factors that affect the discrepancy
   between the log(O/H) abundances derived from
   EW $R_{23}$ and flux $R_{23}$. We may then
   find some ways to modify the EW $R_{23}$ method
  in order to obtain almost identical oxygen abundances to the
  flux $R_{23}$
  on the basis of this large set of SDSS galaxies.

\subsection{Factors that affect the $\alpha$ parameter}

In Sect.3, we point out that the $\alpha$ parameter is
the ratio of the intrinsic continua underlying
[O~{\sc ii}]$\lambda $3727 and H$\beta$ and that it is
related to the observed continua and the line
extinction $c$ (see Eq.~(\ref{R23})). In this section, we discuss
the dominant effect on the $\alpha$ parameter.

 Equation~(\ref{R23}) shows $\alpha$ is a function of the dust extinction 
 and stellar populations of the galaxies, so then we have
\begin{eqnarray}
\alpha & = & {{F_{\rm C,[OII]}} \over {F_{\rm C,H\beta}}} 10^{c{(f(\rm [OII])}-f(H\beta))} \nonumber \\
    & = & {{F^0_{\rm C,[OII]}} \over {F^0_{\rm C,H\beta}}} 10^{(c-c^*){(f(\rm
    [OII])}-f(H\beta))},
\end{eqnarray}
where $F^0_{\rm C,[OII]}$ and $F^0_{\rm C,H\beta}$ are the dereddened
continua underlying [O~{\sc ii}] and H$\beta$,
$c$ the dust attenuation on emission-line,
and $c^*$ characterizes the dust attenuation on the
continuum.
Here we assume $c^*$ and $c$ follow the same reddening law.
(Also see KP03)

As a rough estimate, we assume that the sample galaxies follow the same
extinction law as the starburst galaxies studied by Calzetti et al. (1994), who
found that the difference between the optical depths of
the continua underlying
the Balmer lines is about one-half of the difference between
the optical depths
of the Balmer emission lines (their Eq.26).
Thus, we assume $c^*\approx 0.5c$, and then obtain the following equation:
\begin{eqnarray}
\alpha = {{F^0_{\rm C,[OII]}} \over {F^0_{\rm C,H\beta}}} 10^{0.5c{(f(\rm
    [OII])}-f(H\beta))},
\end{eqnarray}
where ${{F^0_{\rm C,[OII]}}/{F^0_{\rm C,H\beta}}}$ characterizes the stellar
populations of the galaxies, and $c$-term characterizes the dust extinction.

It is easy to check the relation between $\alpha$ and the dust extinction
$c$ since
we have obtained both of them for the individual galaxies.
However, we cannot obtain the intrinsic
values of ${{F^0_{\rm C,[OII]}}/{F^0_{\rm C,H\beta}}}$ directly.
Fortunately,
the MPA/JHU group and the SDSS provide $D_n$(4000) parameters and several
photometric magnitudes
for this large sample of galaxies, which can
 characterize the stellar populations of the galaxies.
We check the relations between $\alpha$ and dust extinction $c$,
 $D_n$(4000), $g-r$ colors for the sample galaxies
 in the next three sections.

\subsection{The relation between $\alpha$ and dust extinction}

\begin{figure}
\centering
\input epsf
\epsfverbosetrue
\epsfxsize 7.2cm
\includegraphics[bb=29 38 567 390,width=7.2cm,clip]{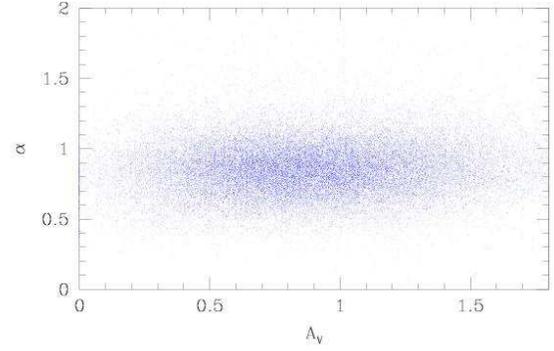}
\caption{ Relation of $\alpha$ parameter and dust extinction $A_V$ for the sample
      galaxies.
       }
\label{figAva}
\end{figure}

 The [O~{\sc ii}]$\lambda $3727 is bluer in wavelength
and is affected more strongly by dust extinction than  H$\beta$,
thus the $\alpha$ parameter may be
correlated with dust extinction $A_V$.
Figure~\ref{figAva} shows the $\alpha$ parameter as a function of dust extinction
$A_V$(=${{cR_V}/{1.47}}$, Seaton (1979), $R_V$=3.1).
The derived $A_V$ is from 0 to 2.7, with 
the median and average values of 
0.87 and 
0.89,
respectively.
It shows that there is no clear correlation between $\alpha$ and $A_V$
for these SDSS sample galaxies, and
the linear least-square fit is 
$\alpha=0.056A_V+0.821$ 
with a very slight slope.
This means that the
differences in the continua underlying [O~{\sc ii}]3727
and H$\beta$ are not affected much by dust extinction.
The reasons may be that the related two lines
are not very far away at wavelength and that the dust
extinction coefficients of these SDSS star-forming galaxies are not so
large.

\subsection{The corrected EW $R_{23}$ method using $D_n$(4000) index}

The break occurring at 4000\AA~ is the strongest discontinuity in the
optical spectrum of a galaxy, and it arises because of the accumulation of a
large number of spectral lines in a narrow wavelength region. The main
contribution to the opacity comes from ionized metals. In hot stars, the
elements are multiply ionized and the opacity decreases, so the 4000-\AA~
break will be small for young stellar populations and large for old,
metal-rich galaxies (Kauffmann et al. 2003).
  Kauffmann et al. (2003)
 adopt the narrow definition from Balogh et al. (1999) and denote
 this index as $D_n$(4000) for the SDSS galaxies. It is the ratio of
 the average flux density $F_\nu$ in the bands 3850-3950 and
 4000-4100 $\AA$, 100\AA~ narrower than the definition by Bruzual (1983).
 The parameter $D_n$(4000) is one of the main ones used by Kauffmann et al. (2003)
to trace the stellar formation history of the SDSS galaxies, which
shows a monotonic increase after
the instantaneous burst of star formation.

\begin{figure}
\centering
\input epsf
\epsfverbosetrue
\epsfxsize 7.2cm
\includegraphics[bb=111 275 253 506,width=7.2cm,clip]{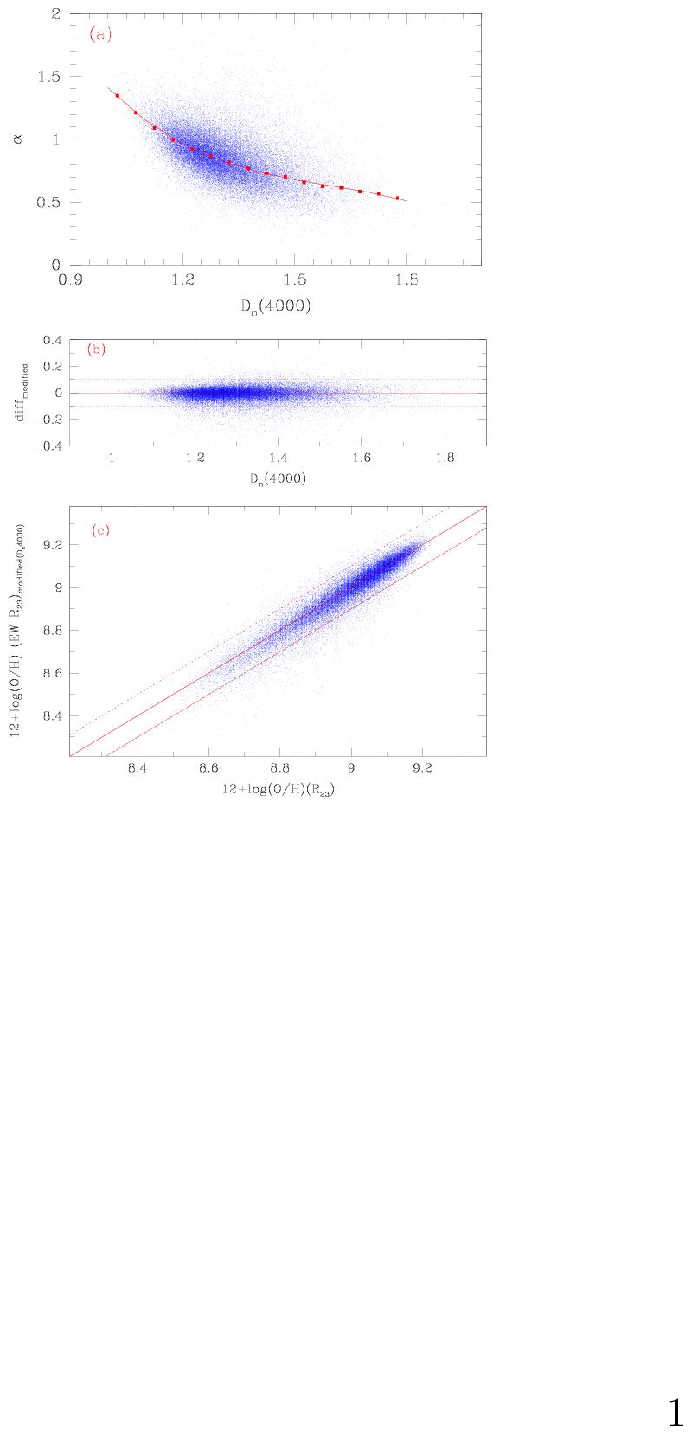}
\caption{ {\bf (a).} Relation of the $\alpha$ and D$_n$(4000) parameters for the sample
      galaxies. The big squares are the 16 median-value points in the
      bins of 0.05 in D$_n$(4000) from 1.0 to 1.8.
      The solid line represents the third-order
      polynomial fit for these median-value points, and it is given as Eq.~(\ref{ecor}).
       {\bf (b).} Comparison between the metallicities derived from the
      $\alpha$-modified EW\,$R_{23}$ by using D$_n$(4000) following Eq.~(\ref{ecor})
      (marked as the subscript ``modified").
       The $diff$ refers to the difference between
       the oxygen abundances from the modified EW\,$R_{23}$
       and flux $R_{23}$.
      {\bf (c).} The direct comparison between the two
      oxygen abundances derived from the modified EW\,$R_{23}$ and flux $R_{23}$
      (extinction-corrected).
      In (b) and (c), the solid lines are the equal-value lines, and
      the dashed lines show the 0.1\,dex discrepancy.
       }
\label{fig3}
\end{figure}

 We plot the relations of $\alpha$ against $D_n$(4000) for the
 sample galaxies in Fig.~\ref{fig3}a, which clearly shows
 a correlation.
  A  third-order polynomial fit for this relation is obtained
  by fitting the 16 median-value points in bins of 0.05
  in $D_n$(4000) from 1 to 1.8, and given as:
\begin{eqnarray}
 \alpha & = & {\rm 10.88} - {\rm 18.31}~ x + {\rm 11.18}~ x^2  - {\rm 2.34} ~x^3,
 \label{ecor}
\end{eqnarray}
 with a standard error of 0.164, where $x=D_n(4000)$. 
  This relation of $\alpha$ vs. $D_n$(4000) could be used to correct
  the EW\,$R_{23}$ and then to obtain the consistent
 metallicities of galaxies with the flux $R_{23}$.
  We propose using
  $R_{23}$(EW)$_{\rm corrected}$= $\alpha$$\times$EW([O~{\sc ii}])/EW(H$\beta$) +
  EW([O~{\sc iii}])/EW(H$\beta$) to then estimate the metallicities of galaxies.

  Figure~\ref{fig3}b shows the consistency of the metallicity estimates
  from the corrected EW\,$R_{23}$ and flux $R_{23}$
  after we apply the correction of  $D_n$(4000) for $\alpha$.
  Figure~\ref{fig3}c shows more direct
  comparisons for the two O/H estimates with such a correction.
  Both of them show that the two derived abundances are very consistent.
 The median and mean discrepancies between
 log(O/H)$_{EW~R_{23}}$ and log(O/H)$_{R_{23}}$ now decrease
 to about  
 $-$0.005\,dex (with a scatter of 
 0.024\,dex), and 
 $-$0.010\,dex (with a scatter of 
 0.054\,dex), respectively. 
 Then the corrected EW\,$R_{23}$ method provides log(O/H) abundances
 similar to those of the extinction-corrected flux $R_{23}$ method within an
 accuracy of $\pm$0.1\,dex for
 $>$94\% of the galaxies.

 Unfortunately this correction is difficult to handle with non-calibrated
spectra, simply because reliable measurements for $D_n$(4000) require
flux-calibrated spectroscopy. Since $D_n$(4000) amplitude depends on the
stellar population, age, and metallicity, it also correlates with colors. In
the following we aim at generating a correction usable by a large community,
using a non-calibrated spectrum and one color as input
(e.g. $g-r$ for the SDSS data).

\subsection{The corrected EW $R_{23}$ method using $g-r$ color}

 Colors can provide important information for the stellar populations
 of the galaxies and can be obtained directly from the photometric observations.
 SDSS has made the $u, g, r, i, z$ band magnitudes of the galaxies available publicly.
 Since the $u$ magnitude has large uncertainty
 (20 per cent; Kauffmann et al. 2003)
 and the images in $u-$ and $z-$band are relatively shallow,
whereas $i-$band image may suffer from the `red halo' effect
(Michard 2002; Wu et al. 2005), we
  use $g-r$ color here to study such corrections for $\alpha$. The $g$ and $r$
  magnitudes can be converted to other band magnitudes, e.g. $B$, $V$,
   following some conversions,
  for example, Smith et al. (2002) and Jordi et al. (2006).

\begin{figure}
\centering
\input epsf
\epsfverbosetrue
\epsfxsize 7.2cm
\includegraphics[bb=111 275 253 506,width=7.2cm,clip]{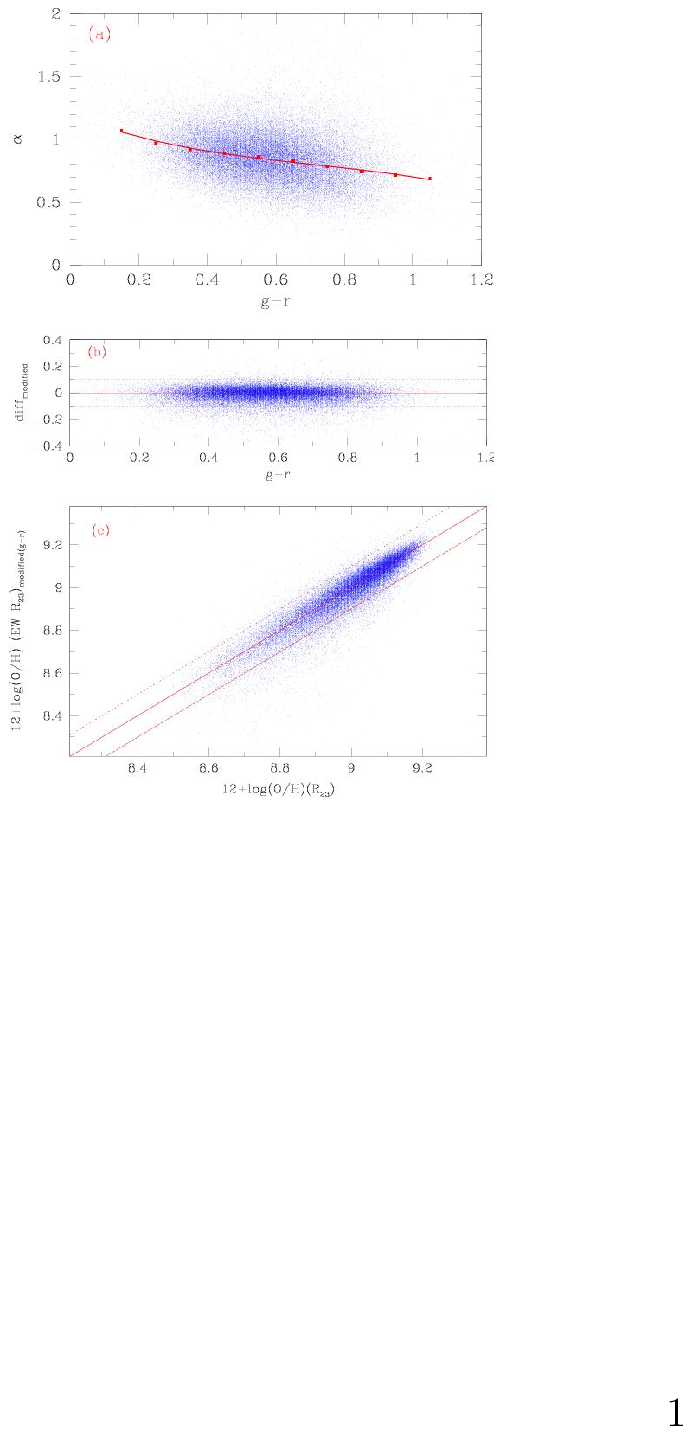}
\caption{ {\bf (a).} Relation of the $\alpha$ and $g-r$ colors for the sample
      galaxies. The big squares refer to the 10 median-value points in the
      bins of 0.1 in $g-r$ from 0.1 to 1.1.
      The solid line represents the third-order
      polynomial fit for these median-value points,
      and it is given by Eq.~(\ref{ecor1}).
       {\bf (b).} Comparison between the metallicities derived from the
       $\alpha$-modified EW\,$R_{23}$ by using $g-r$ colors following
       Eq.~(\ref{ecor1})
      (marked as the subscript ``modified").
      The ``$diff$" is as a function of $g-r$ colors.
      {\bf (c).} The direct comparison between the two
      oxygen abundances from the $\alpha$-modified EW\,$R_{23}$ and the flux $R_{23}$
      (extinction-corrected). The solid lines are the equal-value lines, and
      the dashed lines show the 0.1\,dex discrepancy
      from the equal-value lines in (b) and (c).
       }
\label{fig4}
\end{figure}

 Figure~\ref{fig4}a shows the correlation between the $g-r$
  colors (in Petrosian magnitudes) and $\alpha$ for the sample galaxies.
 The basic trend shows that the redder galaxies
 have relatively lower $\alpha$ values,
 corresponding to the larger differences
 between the underlying continua of [O~{\sc ii}] and H$\beta$.
 The third-order polynomial fits for these correlations were obtained
 and given as
 \begin{eqnarray}
 \alpha & = & {\rm 1.20} - {\rm 1.11}~ x + {\rm 1.15}~ x^2  - {\rm 0.53} ~x^3,
 \label{ecor1}
\end{eqnarray}
 with a standard error of 0.186, 
 where $x$ refers to $g-r$ color.
  Then we propose using
  $R_{23}$(EW)$_{\rm corrected}$= $\alpha$$\times$ EW([O~{\sc ii}])/EW(H$\beta$) +
  EW([O~{\sc iii}])/EW(H$\beta$) to estimate the metallicities of galaxies.

  Figure~\ref{fig4}b shows the consistency of the metallicity estimates
  from the corrected EW\,$R_{23}$ and flux $R_{23}$
  after we apply the correction from  $g-r$ color for $\alpha$.
  Figure~\ref{fig4}c shows the
  comparison between the two O/H estimates with this correction more directly.
  They show that the median and mean discrepancies between
 log(O/H)$_{EW~R_{23}}$ and log(O/H)$_{R_{23}}$ now decrease
 to about 
 $-$0.004\,dex  (a scatter of 
 0.030\,dex) and 
 $-$0.012\,dex (a scatter of 
 0.062\,dex), respectively. 
 Then the two oxygen abundance estimates are almost identical to each other.
 Namely, the corrected EW\,$R_{23}$ method provides log(O/H) abundances
 similar to those of the extinction-corrected flux $R_{23}$ method within an
 accuracy of $\pm$0.1\,dex for
 $>$92\% of the galaxies.

\subsection{Correction for the EW $R_{23}$ method by a constant $\alpha$=0.85}

  The calculated median $\alpha$ value of this large sample of
  SDSS local star-forming
  galaxies is about 0.85 (the average value is $\sim$0.86).
  We may also use this
  constant $\alpha$-factor to modify the EW\,$R_{23}$.
  However, this constant correction is only useful for estimating the global
  metallicity distribution of a large dataset, for example, from the database of
  a survey. As for the individual galaxy,
  this may reversely enlarge the uncertainty for some of them, for example,
  the object with about EW(H$\beta$)$\sim$10\AA (see Fig.~\ref{fig1}a).
  Indeed, the correction on EW\,$R_{23}$ for the individual galaxies
  correlates tightly 
  with their stellar populations and star formation histories.

\section{Discussions about the boundary of the upper branch of 12+log(O/H)}
\label{sec60}

In this work, we adopt 12+log(O/H)$\sim$8.5 as the low boundary of 
the upper branch of oxygen abundances
from the log$R_{23}$ calibration.
The main reason is that the $R_{23}$ calibration 
used (taken from Tremonti et al. 2004) 
is valid above this metallicity. 
In Sect.~\ref{sec60}.1, we present more observational data
with O/H abundances derived from electron temperature $T_e$, 
especially from the recent SDSS,
and some photoionization models to further identify
the reason we adopt 12+log(O/H)$\sim$8.5
as the boundary of the upper branch of oxygen abundances
to compare the EW $R_{23}$ and flux $R_{23}$ methods.
However,
this boundary value is a bit higher than used by
some other researchers, e.g., Pilyugin (2000, 2001a,b) and
Pilyugin \& Thuan (2005), who use 12+log(O/H)$\sim$8.2
on the basis of a sample of H~{\sc ii} regions having their
log(O/H) abundances estimated from $T_e$.
Therefore, we extend the boundary to compare the EW~$R_{23}$ and 
flux $R_{23}$ methods and check whether our relations are valid or not
in the region 
of 12+log(O/H)$_{T_e}$$\sim$ 8.2-8.5, which will
be presented in Sect.~\ref{sec60}.2. 

\subsection{The observational data with (O/H)$_{T_e}$ and the photoionization models}

\begin{figure}
\centering
\input epsf
\epsfverbosetrue
\epsfxsize 7.2cm
\includegraphics[bb=109 349 252 631,width=7.2cm,clip]{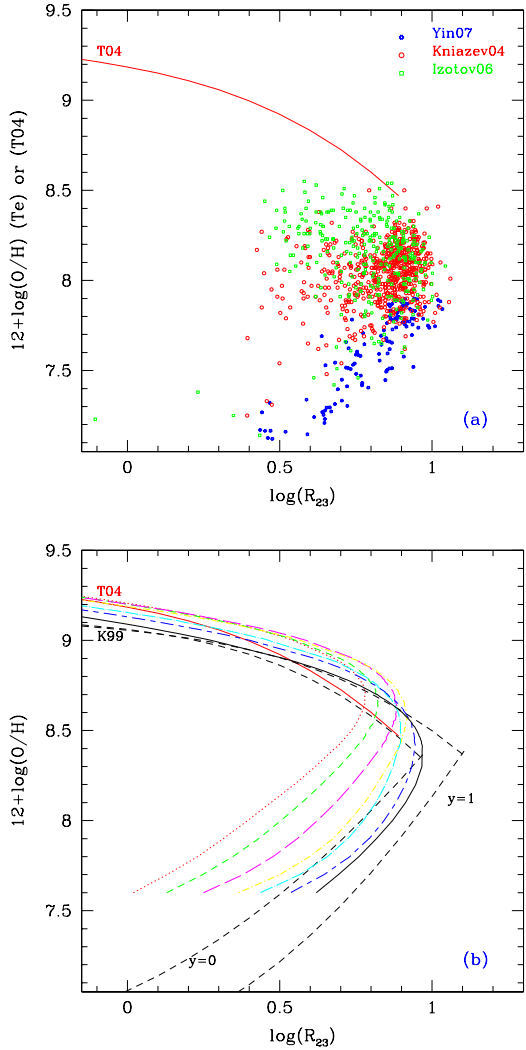}
\caption{  {\bf (a).} The $T_e$-based abundances and log$R_{23}$ of
the galaxies and H~{\sc ii} regions taken from
Yin et al. (2007, Yin07), Kniazev et al. (2004, Kniazev04), 
and Izotov et al. (2006, Izotov06).
Associated with the calibration from SDSS-DR2 observations
given by Tremonti et al. (2004; the thick solid line, marked by 'T04' ).
{\bf (b).} The predictions of photoionization models for the relations
of 12+log(O/H) vs. log($R_{23}$):
the two dashed lines marked with $y$ values are taken from
Kobulnicky et al. (1999, K99), which are  the analytical formulas
for McGaugh (1991);
the seven thin lines down to 12+log(O/H)$\sim$7.6
are taken from Kewley \& Dopita (2002) with different
ionization parameter $q$ values. The thick solid line is the same as (a).
Both (a) and (b) show that it is very difficult to
derive the reliable O/H abundance from the $R_{23}$ parameter
for the cases of 12+log(O/H)$<$8.5. (See the online color versions
of the plots for the discrepancies of the data and models.)
        } 
\label{figte}
\end{figure}

In Fig.~\ref{figte}a we plot the (O/H)$_{T_e}$-$R_{23}$ relations for the sample
galaxies having $T_e$-based O/H abundances, which are
taken from Kniazev et al. (2004, 624 samples), Izotov et al. 
(2006, 409 samples), and Yin et al. (2007, 695 samples).
Figure~\ref{figte}b presents the predictions of photoionization models for the relations
of 12+log(O/H) vs. log$R_{23}$ taken from 
Kewley \& Dopita (2002) and
Kobulnicky et al. (1999, K99), which was obtained by analyzing
those of  McGaugh (1991).
 
Figure~\ref{figte}a and Fig.~\ref{figte}b show that
both the large sample of observational data and photoionization models 
confirm that it is very difficult to
derive a reliable O/H abundance from the $R_{23}$ parameter
for the region of 12+log(O/H)$<$8.5 because of 
the large scatter of the data points there,
the weak dependence of O/H on $R_{23}$ (down to 12+log(O/H)$<$7.9), 
and the strong effects of ionization parameters. 
Yin et al. (2007) has used their Fig.3b to show 
the large discrepancy, up to 0.4\,dex, 
of the two sets of log(O/H) estimates derived from the upper branch and
lower branch of $R_{23}$
calibrations for the samples within 7.9$<$12+log(O/H)$<$8.4.
Stasinska (2002) discusses the weak dependence of (O/H) on $R_{23}$ 
in the transition region from nebular physics.

Considering the discussions above, and also to avoid that some
galaxies included may lie in the lower branch of the $R_{23}$-(O/H)
relation, we select the galaxies having 12+log(O/H)$>$8.5
to check the EW $R_{23}$ method in this study. 
However, it could be useful to check this and 
the validity of our relations in
an extended upper branch range, 12+log(O/H)$\sim$8.2-8.5,
as used in some studies. 

\subsection{Checking the extended upper-branch range
 of 12+log(O/H)$\sim$8.2-8.5}

 We select a subsample from Fig.~\ref{figte}a to check the 
 situation in the extended range of upper branch, 12+log(O/H)$\sim$8.2-8.5.
 To be consistent with other parts of this work, we select this subsample 
 based on the SDSS-DR2 catalog.
 Finally, 37 independent objects with metallicities of
 8.2$<$12+log(O/H)$_{T_e}$$<$8.5 are selected by  
 cross-correlating the DR2 catalog with the lists of 
 Kniazev et al. (2004, from DR1), 
 Izotov et al. (2006, from DR3), and Yin et al. (2007, from DR4).
 Then we use the EW and flux measurements of the related emission lines 
 provided by the MPA/JHU group to
 estimate their EW~$R_{23}$ and flux $R_{23}$, hence, the resulted abundances.

 These 37 galaxies show a discrepancy between log(EW$R_{23}$) and log($R_{23}$)
 (as Fig.~2b) within a range of -0.2 to 0, which may
 result in overestimated log(O/H) abundances of about 0-0.2\,dex 
 by the EW$R_{23}$ as shown in Fig.~2c.  
 Their $D_n$(4000) values are around
 1.0, with $g-r$ colors around 0.0 (ranging from -0.2 to 0.3), 
 which confirms that they are low-metallicity
 objects that will distribute in the left hand parts in Figs.~4a and 5a.
 Their $\alpha$ values are within 1-2, with the average value about 1.5.
 If we apply the $\alpha$ corrections for their EW~$R_{23}$,
 the correction factors will be about $\alpha$$\sim$ 1.4 and 1.2
 by extrapolating Eqs.~(8) and (9), respectively.

 If we do not consider the discrepancy among the different
 photoionization parameters and the weak dependence of O/H on $R_{23}$ 
 in this metallicity region (shown as Fig.6b) 
 and try to extrapolate the $R_{23}$ 
 calibration of Tremonti et al. (2004) down to
 these objects with 8.2$<$12+log(O/H)$_{T_e}$$<$8.5, 
 then the open squares in Fig.~7a show that
 the $R_{23}$ method will often overestimate the 
 abundances of the galaxies and even underestimate
 the abundances of some galaxies.
 The stars in Fig.7a show that the EW~$R_{23}$ method provides
 a higher log(O/H) abundance than the flux $R_{23}$,
 generally about 0.2\,dex, which is consistent with the discussions above
 and with Fig.2c. 
 Figure~7a also shows that
 the correction by $\alpha$($g-r$) will not improve
 the consistency for these objects much, and the correction by $\alpha$($D_n(4000)$)
 gives more consistent O/H abundances for the objects with
 12+log(O/H)$_{R_{23}}$$>$8.2.  
 We changed to the $R_{23}$ calibration of 
 Zaritsky et al. (1994) (which is an average
 of three previous calibrations) to obtain these comparisons again,
 and find the results are very similar to Fig.7a.

 We also used the $R_{23}$ calibration of Pilyugin (2000) (their Eq.5)
 to do these comparisons, and present the results in Fig.~7b. 
 It shows that mostly the flux $R_{23}$ method provides 
 consistent log(O/H) abundances with the $T_e$ method,
 except for some data points in the left hand part of the data with underestimated
 log(O/H)$_{R_{23}}$ and some reversed ones in the right hand section. 
 It also shows that the corrected EW~$R_{23}$ method by $\alpha$($D_n(4000)$)
 provides more consistent abundances with the flux $R_{23}$
 for the objects with 12+log(O/H)$_{R_{23}}$$>$8.1
 than the uncorrected EW~$R_{23}$ or corrected by $\alpha$($g-r$),
 which are similar to Fig.~7a. 
 These differences among the EW~$R_{23}$ abundances presented in Fig.7
 are not difficult to understand since the average $\alpha$ value
 of these low-metallicity objects is about 1.5, which is higher than
 the correction relations provided, $\sim$1.4 or 1.2.
 
 In short, in the lower metallicity range of  8.2$<$ 12+log(O/H)$_{T_e}$$<$8.5,
 we would not recommend using our relations to correct the
 EW~$R_{23}$ method for the oxygen abundance calibrations, since 
 there are several situations affecting the results, such as
 the big scatter of the data, the
 strong effects of ionization parameters, and the weak dependence of
 O/H on $R_{23}$ in the range.  
  
\begin{figure}
\centering
\input epsf
\epsfverbosetrue
\epsfxsize 7.2cm
\includegraphics[bb=109 349 252 631,width=7.2cm,clip]{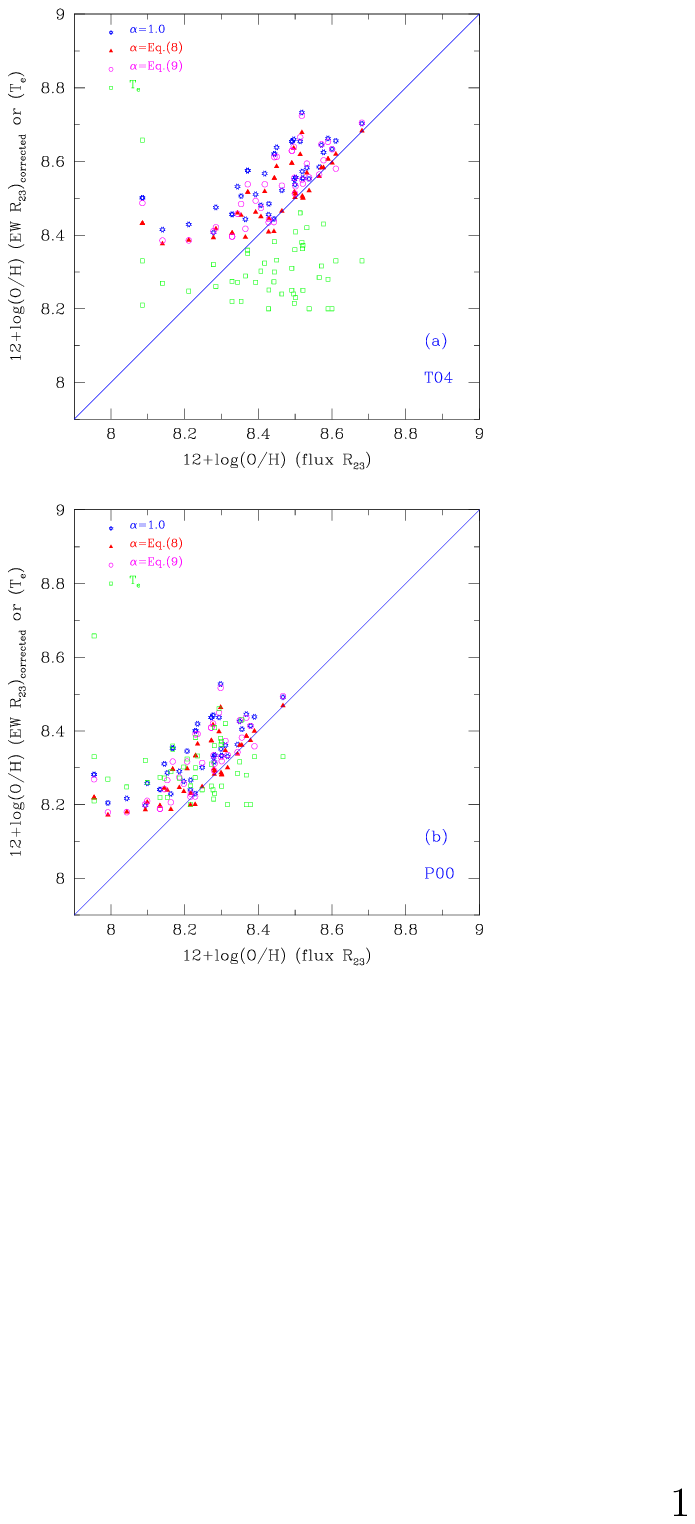}
\caption{ Comparisons between the 
log(O/H) abundances obtained from flux $R_{23}$ and those from 
$T_e$ or from the corrected or uncorrected EW~$R_{23}$ method:
{\bf (a)} using the $R_{23}$ calibration of Tremonti et al. (2004, T04);
{\bf (b)} using the $R_{23}$ calibration of Pilyugin (2000, P00).
The different symbols represent those what obtained from the different methods
marked by the labels in the top left corner.
       } 
\label{figr23c}
\end{figure}

\section{Summary and conclusion}
\label{sec6}

The goal of this paper is to check the reliability of using
EW\,$R_{23}$ (=${\rm EW({\rm [OII]})+EW({\rm [OIII])}\over {EW(H\beta)}}$)
to replace the extinction-corrected flux $R_{23}$
(=${{I({\rm [OII]})+I({\rm [OIII]})}\over {I(H\beta)}}$)
to estimate the metallicities of star-forming galaxies
on the basis of a large sample 
(37,173) of galaxies with 12+log(O/H)$_{R_{23}}$$>$8.5,
selected from the SDSS-DR2.
This replacement is often adopted when there are some problems dealing with
proper flux calibrations for the spectral observations.
This large sample can provide some obvious statistical trends.

The results show that
the logarithm values of EW\,$R_{23}$  and
extinction-corrected flux-$R_{23}$
have a discrepancy from -0.4 to 0.5\,dex, with a median
value of about 
 +0.061\,dex
 and a mean value about 
 +0.069\,dex.
 Thus, the discrepancies between the log(O/H) abundances obtained
 from EW\,$R_{23}$ and those from the flux-$R_{23}$ range from
 -0.5 to 0.2\,dex and have a median value of about 
 -0.041\,dex,
 a mean value of about 
 -0.054\,dex.
 These discrepancies are caused by the
 different continua ($F_{C\lambda}$) underlying the
 emission lines
 [O~{\sc ii}] and H$\beta$ ([O~{\sc iii}], as well). These differences
 are characterized by
 the $\alpha$ parameter as ($F_{C,H\beta}$)/($F_{C,[OII]}$),
 which changes from 0.1 to 2.6, and by a median value
 of 0.85 and a mean value of 0.86.

Then we discuss the factor that affects this discrepancy mostly.
Our large sample of data shows that the $\alpha$ parameter is almost
independent of the dust extinction inside the galaxies
and depends closely on stellar populations of the galaxies, which can be
quantified by the $D_n$(4000) parameters and colors of the galaxies.
Third-order polynomial fits have been obtained for the observed
relations of $\alpha$ versus $D_n$(4000) and $\alpha$ versus $g-r$ colors
for the sample galaxies,
which can be used to modify the EW\,$R_{23}$ method.
After applying this correction by $D_n$(4000),
the median and mean discrepancies between
 log(O/H)$_{EW\,R_{23}}$ and log(O/H)$_{R_{23}}$ decrease
 to about
 $-$0.005\,dex
 and $-$0.010\,dex, respectively.
After applying this correction by $g-r$ colors,
the median and mean discrepancies between
 log(O/H)$_{EW\,R_{23}}$ and log(O/H)$_{R_{23}}$ decrease
 to about 
 $-$0.004\,dex and
 $-$0.012\,dex, respectively.
 The two derived sets of log(O/H) abundances are almost identical now.

 In summary,
 when there are problems
 with flux calibrations of the spectra,
  the EW\,$R_{23}$ could be used roughly to replace the
  extinction-corrected flux $R_{23}$ to estimate
 the metallicities of star-forming galaxies.
  The discrepancy caused
 by this replacement can be from -0.2 to 0.1\,dex generally.
 This factor is consistent with those found by KP03 and Moustakas \& Kennicutt (2006).
 However, this discrepancy could be large for the different individual galaxies,
 from -0.5 to 0.2, if the underlying continua of
 [O~{\sc ii}] and H$\beta$ ([O~{\sc iii}]) are quite different.
 The $D_n$(4000) parameters and colors of the galaxies are very useful for
 correcting the EW\,$R_{23}$ method, which will then greatly
 decrease the discrepancies and result in an almost identical
 oxygen abundance to the flux-$R_{23}$.

Nevertheless, the modified EW\,$R_{23}$
still suffers from the drawback of the ``double-value" of $R_{23}$
for the O/H estimates, and some other
strong-line ratios are also useful for estimating the metallicities of
galaxies then,
such as [N~{\sc ii}]/H$\alpha$,
([O~{\sc iii}]/H$\beta$)/([N~{\sc ii}]/H$\alpha$), [N~{\sc ii}]/[S~{\sc ii}]
and [O~{\sc iii}]/H$\beta$, etc. (Kewley \& Dopita 2002; Pettini \& Pagel 2004;
Liang et al. 2006; Yin et al. 2007).

\section*{Acknowledgments}
 We thank our referee for the valuable comments and suggestions, 
which helped to improve this work. We thank Rob Kennicutt, Lisa
Kewley, Hong Wu, Licai Deng, Bo Zhang, and Hector Flores for 
valuable discussions related to this
study, and thank Ruixiang Chang, Shiying Shen, Caina Hao, Jing Wang,
and Chen Cao for helpful discussions about the SDSS database. We
thank James Wicker for his help in improving the English expression
in the text. This work was supported by the Natural Science
Foundation of China (NSFC) Foundation under No.10403006, 10433010, 10673002, 
10573022, 10333060, and 10521001; and the
National Basic Research Program of China (973 Program) No.2007CB815404.

\end{document}